\documentclass[aps,prl,preprint,groupedaddress]{revtex4}

\usepackage{graphicx}
\usepackage{newtxtext,amsmath}
\usepackage{amssymb}
\usepackage{tabularx}
\usepackage{booktabs}

\begin{document}


\title{Rotationally cold OH$^-$ ions in the cryogenic electrostatic ion-beam storage ring DESIREE}

\author{H.~T.~Schmidt$^1$}
\email[E-mail: ]{schmidt@fysik.su.se}
\author{G.~Eklund$^1$}
\author{K.~C.~Chartkunchand$^1$}
\author{E.~K.~Anderson$^1$}
\author{M.~Kami\'nska$^{1,2}$}
\author{N.~de~Ruette$^1$}
\author{R.~D. Thomas$^1$}
\author{M.~K.~Kristiansson$^1$}
\author{M.~Gatchell$^1$}
\author{P.~Reinhed$^1$}
\author{S.~Ros\'en$^1$}
\author{A.~Simonsson$^1$}
\author{A.~K\"allberg$^1$}
\author{P.~L\"ofgren$^1$}
\author{S.~Mannervik$^1$}
\author{H.~Zettergren$^1$}
\author{H.~Cederquist$^1$}
\affiliation{$^1$Department of Physics, Stockholm University, SE-10691 Stockholm, Sweden}
\affiliation{$^2$Institute of Physics, Jan Kochanowski University, 25-369 Kielce, Poland}



\date{\today}

\begin{abstract}
We apply near-threshold laser photodetachment to characterize the rotational quantum level distribution of OH$^-$ ions stored in the cryogenic ion-beam storage ring, DESIREE, at Stockholm University. We find that the stored ions relax to a rotational temperature of 13.4$\pm$0.2~K with 94.9$\pm$0.3\,\% of the ions in the rotational ground state. This is consistent with the storage ring temperature of 13.5$\pm$0.5~K as measured with eight silicon diodes, but in contrast to all earlier studies in cryogenic traps and rings where the rotational temperatures were always much higher than those of the storage devices at their lowest temperatures. Furthermore, we actively modify the rotational distribution through selective photodetachment to produce an OH$^-$ beam where 99.1$\pm$0.1\,\% of approximately one million stored ions are in the $J$=0 rotational ground state.
We measure the intrinsic lifetime of the $J$=1 rotational level to be 145$\pm$28~s.
\end{abstract}


\maketitle
\section{Introduction}
Any experimental study of molecular ions would benefit tremendously if all the ions were in the same electronic, vibrational, and rotational state.
This is, however, very difficult to achieve as they always get excited when they are charged.
An important step towards better control of this situation was taken in the 1990s with the room-temperature {\it magnetic} ion-storage rings. Beams of molecular ions could then be stored for tens of seconds, which for most infrared-active ions exceed typical timescales for vibrational relaxation.
This was
demonstrated in studies of dissociative recombination of molecular ions with free electrons~\cite{larssonorel_12}.
The infrared-active HD$^+$ molecule was, for example, cooled to the vibrational ground state within hundreds of milliseconds through spontaneous IR-emission~\cite{Forck93}. 
For infrared-inactive molecules like H$_2^+$, active depletion methods using photo-dissociation~\cite{Schmidt96,Andersen97} or inelastic scattering with low-energy electrons~\cite{Krohn00} had to be used in order to reduce excitations. 
The {\it rotational} excitations, however, could not be controlled in these room-temperature devices.
Here the problem was two-fold. First, the characteristic times for rotational relaxation are much longer than vibrational lifetimes~\cite{Amitay}. Second, even when the molecular ions had reached thermal equilibrium with the surroundings, they would occupy a large number of rotational levels.
As a rare exception, rotationally cold H$_3^+$ ions were produced
by a gas expansion technique~\cite{McCall}, but no general method to produce {\it and maintain} a rotationally cold molecular ion beam was available in the large room-temperature magnetic ion-storage rings.

With the introduction of the often much smaller {\it electrostatic} ion-beam storage rings~\cite{ELISA,MINIRING,Schmidt15} and traps~\cite{ZAJFMAN,Benner,CONETRAP}, it has become  feasible to cool the entire mechanical structures to cryogenic temperatures~\cite{HeMINUS,CTF10,SF6CTF}. In this work, we use the Double ElectroStatic Ion Ring ExpEriment (DESIREE)~\cite{Tho11,DESIREE_Comm} at Stockholm University to study the radiative cooling of OH$^-$ at the macroscopic temperature of 13.5$\pm$0.5~K.
Altogether, there are three {\it cryogenic} electrostatic ion-storage rings in operation 
world-wide~\cite{Tho11,DESIREE_Comm,CSR,CSR11,RICE,RICE12}. Of these, DESIREE is the only one with two storage rings allowing for ion-ion merged-beams experiments~\cite{Tho11,DESIREE_Comm}. For these and other types of experiments, it is important to investigate to what extent the internal distribution of excitation energies in the stored ions may be reduced and manipulated.

The OH$^-$ ion is particularly well-suited to probe the relaxation towards a single-quantum-state molecular-ion beam as the separation of the two lowest rotational states is large (4.6448~meV~\cite{OH-SPLIT}).
For this ion, the storage device temperature would need to be below 9.5~K to reach thermal equilibrium with less than 1\,\% of the stored OH$^-$ ions in rotationally excited states.
A way to effectively approach this scenario is thus 
to store the ions in a sufficiently cold environment for sufficiently long times and to avoid any other heating processes such as collisions with the residual gas. 
In an earlier attempt to produce a pure $J$=0 ensemble, Otto~{\it et al.} trapped OH$^-$ ions in a 22-pole radiofrequency buffer gas trap~\cite{Gerlich} at temperatures between 10 and 300~K~\cite{Otto13}. They studied laser photodetachment near threshold and extracted the distribution on the rotational OH$^-$ levels~\cite{Otto13}. For temperatures around 50~K, they obtained a distribution consistent with the macroscopic temperature of their trap and He buffer gas. For the lowest temperatures, however, the OH$^-$ rotational temperature always remained higher than 20~K for reasons that are not understood~\cite{Otto13,OHColl}.

At the Cryogenic Storage Ring (CSR) in Heidelberg~\cite{CSR,CSR11}, a similar technique was used to extract the rotational temperature of CH$^+$ ions by state-selective laser {\it photodissociation}. From that study, it was concluded that the effective thermal radiation field corresponded to a temperature of about 20~K while the average temperature of the CSR itself was below 10~K~\cite{CHCSR16}.
Very recently, the Heidelberg group measured the rotational distribution of OH$^-$ ions stored in CSR by the near-threshold laser photodetachment method \cite{Otto13} and deduced an effective temperature of 15.1$\pm$0.1 K \cite{OHCSR17}.

Hansen {\it et al.} demonstrated a rotational temperature of 7.5~K for a single MgH$^+$ ion in a small radiofrequency trap~\cite{Drewsen}. This {\it single} MgH$^+$ ion was sympathetically cooled translationally by laser-cooled Mg$^+$ and rotationally cooled by collisions with a He buffer gas. Rotational temperatures below 15 K, however, were only demonstrated with less than 10$^3$ ions. 
For greater numbers of ions, trap micromotions led to more violent collisions and much higher rotational temperatures.

In the present work, we use photodetachment thermometry and deduce an effective rotational temperature of 13.4$\pm$0.2~K for about one million OH$^-$ ions circulating in DESIREE.

\section{Experiment}
At cryogenic operation, the DESIREE vacuum chamber reaches a lowest temperature of 13.5$\pm$0.5~K as measured by eight Lake Shore DT-470 silicon diodes \cite{LakeShore} attached to the vacuum chamber walls. This chamber is thermally insulated from, and mounted inside, a room temperature outer vacuum chamber~\cite{Tho11}. The $\pm$0.5~K error estimate reflects the variation among the eight sensors and the variation in time between the different runs in the present study. At this low temperature, we reach a residual-gas density of only about 10$^4$ hydrogen molecules per~cm$^{3}$. This is evidenced by the long storage times and slow spontaneous decay processes studied in DESIREE with atomic~\cite{Erik,Magda16,KC16} and cluster anions~\cite{Hansen17}. In the current experiment, OH$^-$ ions are produced in a cesium-sputter ion source \cite{SNICS} where a small amount of water vapor flows through a thin channel in a copper/titanium cathode.
The ions are accelerated to 10~keV, then injected and stored in DESIREE. 
\begin{figure}
	\centering
		\includegraphics[width=0.48\textwidth]{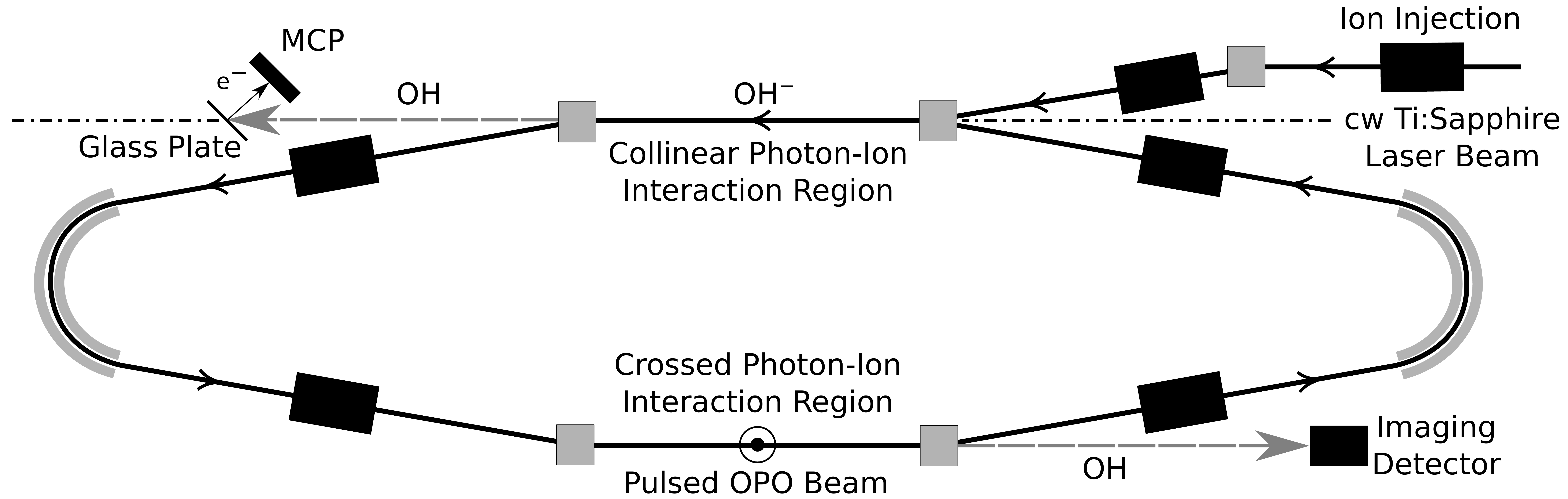}
	\caption{One of the DESIREE storage rings. Neutral OH molecules formed in either straight section by laser photodetachment are detected in their forward directions. In measurement Method\,I, a collinear cw Ti:Sapphire laser beam on the ion-injection side is applied to probe the populations in the rotational levels as functions of time. In Method\,II, the populations of the OH$^-$ rotational levels are probed by a crossed beam from a pulsed, tunable OPO while the cw laser can be used to deplete the $J$$\ge$1 states (see the main text).}
	\label{fig:DESIREE_OH}
\end{figure}

Two laser systems are used to probe the populations of individual OH$^-$ rotational levels in the stored ion beam using two different measurement methods -- Methods I and II (see FIG.~\ref{fig:DESIREE_OH}).
In Method I, a 100 mW cw single-mode beam from a tunable Ti:Sapphire laser is overlapped with the stored ions along the straight section on the ion-injection side. The neutral OH molecules from photodetchment are then detected through the emission of secondary electrons when they hit a glass plate. The plate is covered by thin layers of titanium and gold to prevent charge build-up and to optimize secondary-electron emission. The electrons are accelerated by an electric field and detected by a microchannel plate (MCP) detector.
In Method II (see FIG.~\ref{fig:DESIREE_OH}), the ion beam is intersected transversely in the other straight section by a beam from a widely tunable optical parametric oscillator (OPO) with pulse duration, pulse energy, and repetition rate of 5~ns, $\sim$2~mJ, and 10~Hz, respectively. The OPO linewidth is 5~cm$^{-1}$ ($\sim$0.6~meV).
Neutrals produced in photodetachment by the OPO are detected by an imaging detector consisting of a triple-stack MCP assembly and a phosphor screen anode viewed by a CMOS camera and a photomultiplier tube~\cite{rosen07}. Method II gives less total signal than Method I, but a coincidence requirement with the 5 ns wide OPO pulses makes it essentially background free~\cite{KC16}. 

The OH$^-$ ion-beam storage lifetime was measured to be 603$\pm$27~s by recording the time dependence of the photodetachment signal with the OPO set to $\lambda$=677~nm without using the cw laser. At this wavelength, the photon energy exceeds the electron affinity of OH (1.827 6488(11)~eV~\cite{Goldfarb}) so that all OH$^-$ ions in the ring can be photodetached. For this measurement, the photodetachment signal rate was limited to about 1\,s$^{-1}$ with 10$^6$ stored ions, making the laser-induced rates negligibly small in comparison with other ion-loss rates.
With our long ion-beam storage time, the main question is if we will reach {\it thermal} equilibrium at 13.5$\pm$0.5~K or if residual-gas collisions or absorption of higher-energy photons from the outside will shift the distribution to higher temperatures.

\section{Spontaneous rotational cooling}
\begin{figure}
	\centering
		\includegraphics[width=0.49\textwidth]{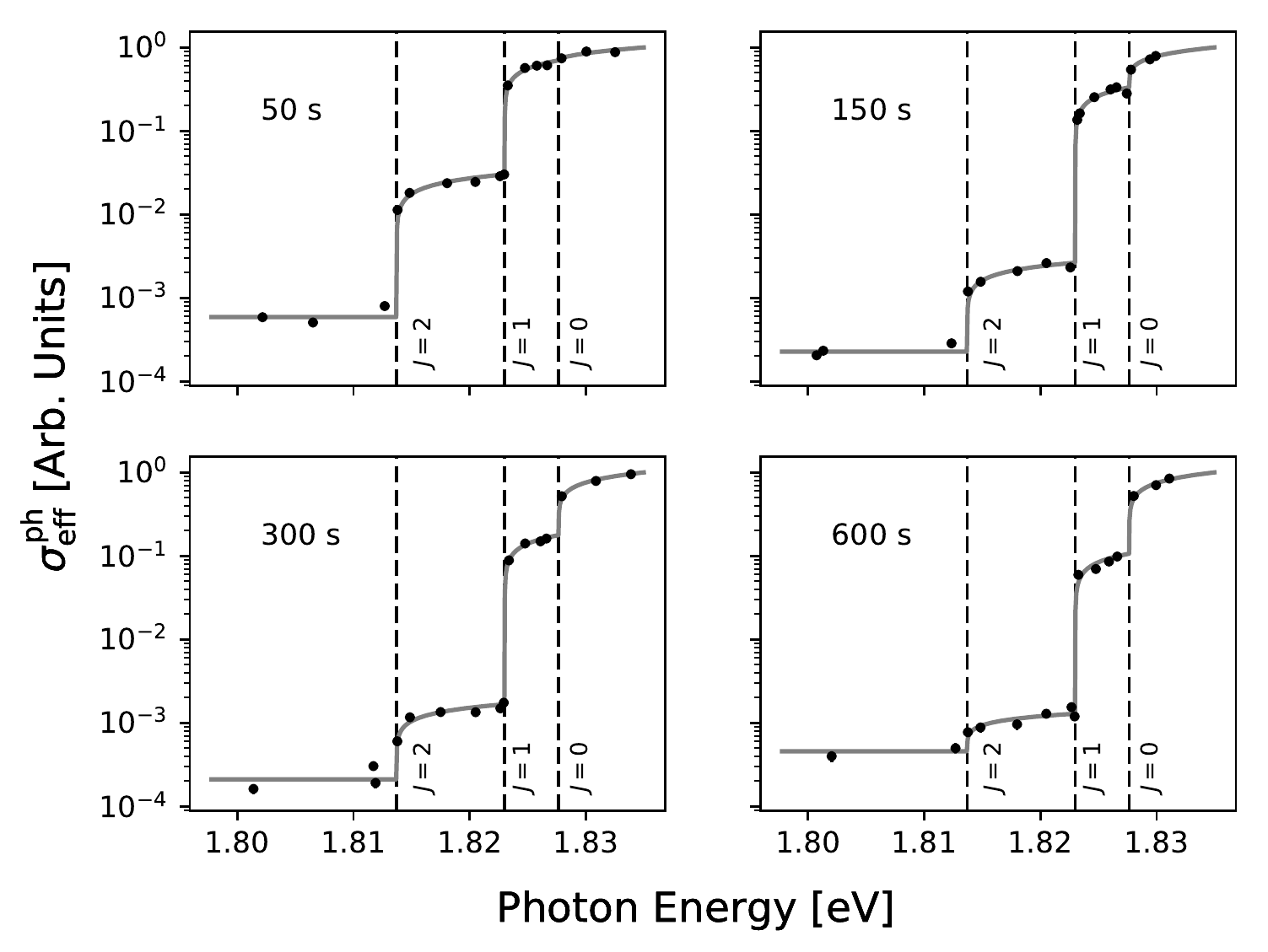}
	\caption{Measured effective relative  photodetachment cross-sections close to the $J$=0, $J$=1, and $J$=2 detachment thresholds for different delays between ion injection and probing with the cw laser (Method I). The full curves are fits to Eq.~\ref{fitfunc}. Statistical errors are smaller than the symbols unless they are shown.
	}
	\label{fig:thresholds}
\end{figure}
In FIG.~\ref{fig:thresholds}, we show relative effective photodetachment cross-sections close to the thresholds for the $J$=0, $J$=1, and $J$=2 rotational states of OH$^-$. These data have been recorded with Method I (see FIG. \ref{fig:DESIREE_OH}).
The ions were stored and left undisturbed in the
storage ring for the times indicated in the panels of FIG~\ref{fig:thresholds}. After these given storage times, a shutter to the cw Ti:Sapphire laser beam was opened and the yields of neutral products from photodetachment were recorded by the glass plate/MCP assembly.
This was repeated for a number of wavelengths to yield the relative effective photodetachment cross-sections, $\sigma_{\mathrm{eff}}^{\mathrm{ph}}$, shown in FIG.~\ref{fig:thresholds}. For each panel of FIG.~\ref{fig:thresholds}, the data were normalized to the count rate of neutrals from residual-gas collisions recorded prior to the opening of the shutter for the laser beam. This small background contribution was subtracted before normalization.

Already after 50 s storage (upper left panel in FIG.~\ref{fig:thresholds}), we only find significant OH$^-$ signals for $J$$\le$2. We ascribe the weak, storage-time independent signal for photon energies below the $J$=2 threshold to a small beam impurity of $^{17}$O$^-$.
Slow variations in ion source conditions led to the somewhat different (low) $^{17}$O$^-$ levels as can be seen in the different panels of FIG.~\ref{fig:thresholds}.
To extract the populations of the rotational levels from the effective relative cross-section data in FIG.~\ref{fig:thresholds}, we fit the data to
\begin{equation}
\sigma_{\mathrm{eff}}^{\mathrm{ph}}(h\nu) = \sum_{J=0}^{2}K_\mathrm{J}(h\nu-\epsilon_\mathrm{J})^p+C\mathrm{,}
\label{fitfunc}
\end{equation}
in which $J$ are the rotational quantum numbers of the OH$^-$ ion, $\epsilon_\mathrm{J}$ are the corresponding energy thresholds \cite{Goldfarb}, $p$=0.28 is an empirically determined exponent~\cite{p_ref}, and $C$ are constants representing the small $^{17}$O$^-$ contributions. $K_\mathrm{J}$ are the fit parameters related to the relative populations of $J$-levels, $P(J)$, through
\begin{equation}
K_\mathrm{J} \propto P(J)I_\mathrm{J}\mathrm{,}
\label{KJ}
\end{equation}
in which $I_\mathrm{J}$ are $J$-dependent intensity factors extracted from Ref.~\cite{Goldfarb}.
The relative populations, $P(J)$, from the present fits are given in TABLE~\ref{tab:temperatures}.



\begin{table}
\begin{center}
\begin{tabular}{c|c|c|c|c}
  $t$[s] & $P$($J$=0) & $P$($J$=1) & $P$($J$=2) & $T$[K]\\
  \midrule
  50 & \,20.3$\pm$9.9 & \,76.0$\pm$9.4 & \,3.8$\pm$0.5 & --\\
  100 & \,62.2$\pm$3.1 & \,37.3$\pm$3.0 & \,0.45$\pm$0.04 & \,32.5$\pm$2.2\\
  150 & \,76.3$\pm$1.5 & \,23.5$\pm$1.5 & \,0.19$\pm$0.02 & \,23.6$\pm0.8$\\
  300 & \,89.1$\pm$0.6 & \,10.8$\pm$0.6 & \,0.100$\pm$0.006 & \,16.8$\pm$0.4\\
  600 & \,93.9$\pm$0.2 & \,6.1$\pm$0.2 & \,0.055$\pm$0.005 & \,14.1$\pm$0.2\\
  \bottomrule
\end{tabular}
\end{center}
\caption{Relative populations $P(J)$ (in~\%) in the lowest three rotational levels after different storage times, $t$(s), in DESIREE. The temperatures are the ones that would give the measured $P$($J$=0) in thermal equilibrium. No temperature value is given for the distribution at 50~s, which is too far from thermal equilibrium for an assignment to be meaningful.} \protect\label{tab:temperatures}
\end{table}

The dominance of $J$=1 after 50~s
of storage is due to the $J$$\ge$2 excitations having lifetimes much shorter than our {\it effective} $J$=1 lifetime of
135$\pm$26~s
(as will be discussed below, this is {\it not} the {\it intrinsic} lifetime as the decay is affected by the weak blackbody radiation field in the ring).
The populations in excited levels decrease during storage, and after 300~s less than 0.1\,\%
of the stored OH$^-$ ions occupy levels with $J$$\textgreater$1 (see TABLE \ref{tab:temperatures} and FIG. \ref{fig:thresholds}).
The temperatures in TABLE \ref{tab:temperatures} are based on the fractions of the beam occupying the $J$=0 level as---for the shorter storage times---the distributions are far from thermal equilibrium.

Only 0.055$\pm$0.005\,\% of the ions are in $J$=2 after 600~s,
and thus the role of any excitation besides that from blackbody radiation of the cold ring itself must be very small.
At $t$=600~s, the OH$^-$ ions are well characterized by a temperature of 14.1$\pm$0.2~K, close to the 13.5$\pm$0.5~K temperature of the storage ring and vacuum chamber. Below, we will use measurement Method II (see Fig.~\ref{fig:DESIREE_OH}) and show that the {\it asymptotic} value of the temperature is in fact even lower than this.

\section{Laser-assisted "cooling" and repopulation}
By continuously applying the collinear cw laser at a wavelength where only ions with $J$$\ge$1 can be photodetached, we significantly alter the rotational distribution of the stored ions. 
We use Method II to probe the OH$^-$ rotational distribution as a function of time with the OPO.
The spectral linewidth of the OPO is sufficiently narrow to address the individual rotational levels of OH$^-$.
\begin{figure}
	\centering
 \includegraphics[width=0.49\textwidth]{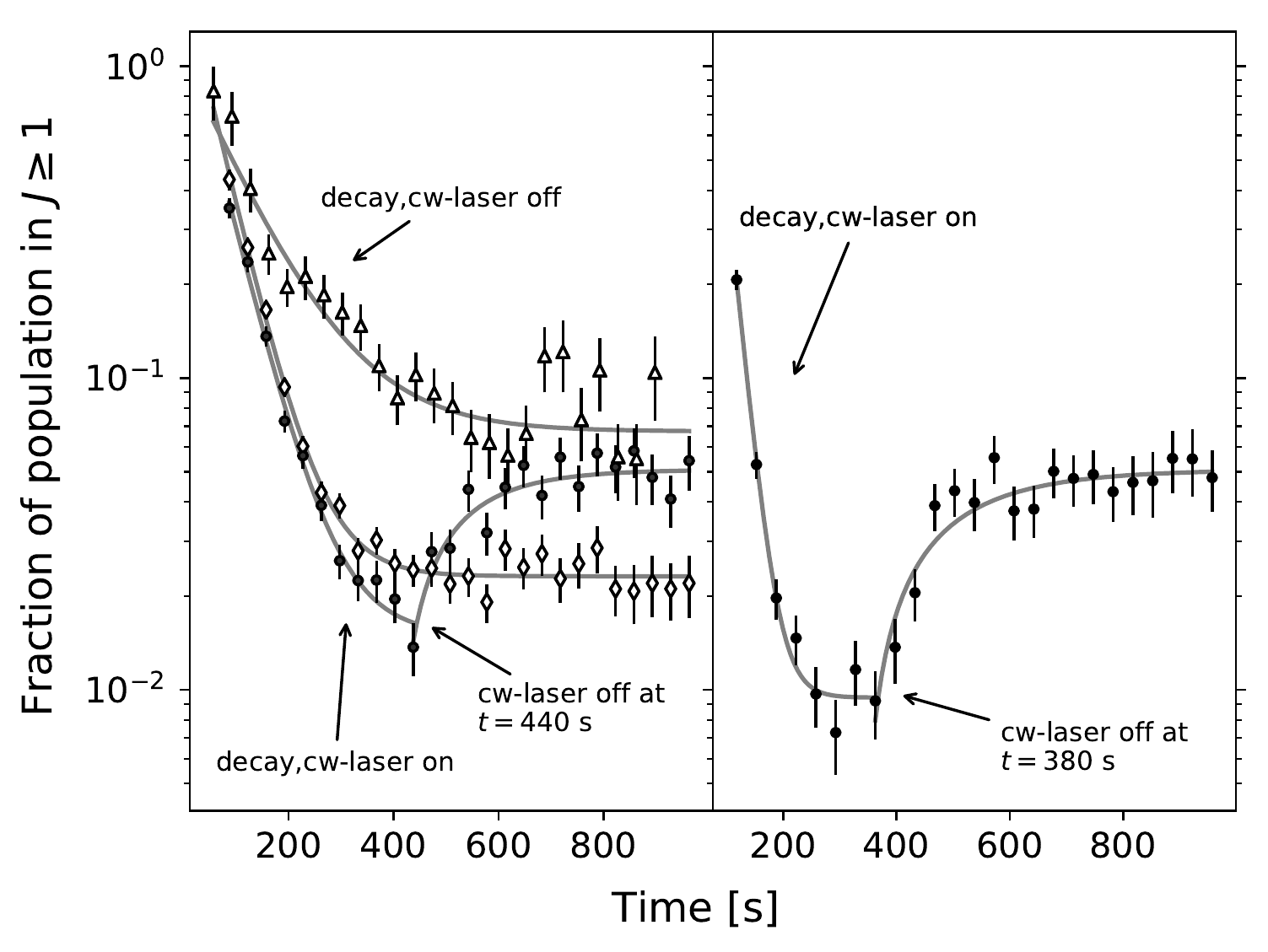}
	\caption{Fractions of the stored ions that are in rotationally excited states ($J$$\ge$1) as functions of time after injection recorded with Method II. These fractions are deduced from the ratios between the photodetachment signals recorded with the OPO at $\lambda$=679~nm ($J$$\ge$1 detachment) and at $\lambda$=677~nm ($J$$\ge$0 detachment). Left panel: No laser depletion (open triangles), depletion laser on at $t$=90~s (open diamonds), depletion laser on at $t$=90~s and off at $t$=440~s (filled circles). Right panel: Depletion laser on at $t$=110~s and off at $t$=380~s (filled circles). The variations of the $J$$\ge$1 fractions found with the depletion laser on (open diamonds in the left panel and filled circles in both panels) reflect the difficulty in reproducing the exact overlap of the cw-laser and stored ion beams between runs.}
	\label{fig:Depletion_All}
\end{figure}

The output of the OPO was set to $\lambda$=677~nm 
or 679~nm, where OH$^-$ ions with $J$$\ge$0 or $J$$\ge$1 contribute to the photodetachment signal, respectively.
From the ratio of these signal rates (normalized to the ion and laser intensities) and the knowledge of the dependence of the effective photodetachment cross-section on the wavelength (Eq.~\ref{fitfunc} and Eq.~\ref{KJ}), the fraction of the ion population in excited rotational levels, $J$$\ge$1, is deduced. The open triangles in the left panel of FIG.~\ref{fig:Depletion_All} show this fraction as a function of time after injection into the ring, when the ions are stored and left alone to decay in the 13.5$\pm$0.5~K blackbody radiation field. For the open diamonds, again deduced from two measurement series with the OPO at 677~nm and 679~nm, we have reduced the degree of excitation in the circulating OH$^-$ beam by applying the cw laser continuously at 679~nm where all
$J$$\ge$1 ions (as well as any contaminant atomic $^{17}$O$^-$ ions) can be photodetached. By this method we shift the
equilibrium rotational distribution closer to the ideal pure OH$^-$ ($J$=0) beam. For this particular data set, we reached a $J$$\ge$1 fraction of 2.31$\pm$0.06\,\%. For the filled circles in the left panel of FIG.~\ref{fig:Depletion_All}, we first reduce the $J$$\ge$1 population (and eliminate any $^{17}$O$^-$ contribution) by the depletion laser and then block this laser at $t$=440~s after injection, whereupon the $J$=1 population {\it increases} toward an equilibrium value of 5.08$\pm$0.37\,\% as the ion ensemble is re-heated in the blackbody radiation field.
This asymptotic value is lower than that obtained without the depletion laser (open triangles in the left panel of FIG.~\ref{fig:Depletion_All}---asymptotic $J$$\ge$1-fraction: 6.74$\pm$0.63 \%).
We ascribe this difference to the permanent removal of $^{17}$O$^-$-contaminations by the depletion laser. The full curves in FIG.~\ref{fig:Depletion_All} are fits to the sum of an exponential function and a constant, in which the constants reflect the asymptotic values
and the time constants of the exponentials are {\it effective} lifetimes of the OH$^-$ ($J$=1) with and without the depletion laser.

In the right panel of FIG.~\ref{fig:Depletion_All}, we show data that are equivalent to the filled circles of the left panel but with a better overlap between the ion beam and the depletion laser (turned on at $t$=110~s and off at $t$=380~s). This leads to a faster decrease in the $J$$\ge$1 fraction and a lower equilibrium value of 0.9$\pm$0.1\,\%.
After blocking the depletion laser at $t$=380~s (right panel of FIG.~\ref{fig:Depletion_All}), we observe an increase of the $J$$\ge$1-fraction to 5.08$\pm$0.41\,\%, in agreement with what was found for the filled circles in the left panel.

The weighted average of our two results for the asymptotic $J$$\ge$1-fraction is $F$=5.1$\pm$0.3\,\%.
In thermal equilibrium, the ratio of the $J$=1 and $J$=0 populations is equal to the Boltzmann factor
\begin{equation}
R(T)=3\exp \left( -\frac{\Delta E}{k_{\mathrm{B}}T} \right)\mathrm{,}
\label{Boltzmann}
\end{equation}
in which $k_{\mathrm{B}}$ is the Boltzmann constant and $\Delta$$E$=4.6448~meV the energy splitting between $J$=0 and $J$=1 in OH$^-$~\cite{OH-SPLIT,Goldfarb}. The factor of~3 is the ratio of the statistical weights of these two levels. Our $J$$\ge$1 fraction,  $F$=5.1$\pm$0.3\,\%, corresponds to an equilibrium temperature of 13.4$\pm$0.2~K, in agreement with the macroscopic temperature 13.5$\pm$0.5~K of the storage ring. This shows that the OH$^-$ ions' rotational degree of freedom is in thermal equilibrium with the macroscopic temperature of their surroundings.


We now discuss the time dependence of the populations to deduce the {\it intrinsic} lifetime of the $J$=1 level. 
The time constants extracted from the repopulation curves after blocking the depletion laser (filled circles in both panels of FIG.~\ref{fig:Depletion_All}) agree within their errors, and have a weighted average of $\Gamma_{\mathrm{eff}}$=135$\pm$26~s in which $\Gamma_{\mathrm{eff}}$ is the {\it effective} decay rate of the $J$$\ge$1 {\it fraction}.
Since the repopulation curves are recorded after long enough storage times such that the $J$$\textgreater$1 population is less than 0.1\,\%, we now consider OH$^-$ to be a two-level system with only $J$=0 and $J$=1.
The interaction of such a system with a single-temperature blackbody radiation field gives a non-zero asymptotic $J$=1 fraction and a faster relaxation of this fraction than at $T$=0 K.
Solving the corresponding rate equations gives the effective decay rate, $\Gamma_{\mathrm{eff}}=A_{10}+B_{10}\,\rho^T(\omega)+B_{01}\,\rho^T(\omega)$ using the Einstein coefficients for spontaneous emission, $A_{10}$, stimulated emission, $B_{10}$, and absorption, $B_{01}$.
Here, $\rho^T(\omega)$ is the spectral density at the frequency of the $J$=0$\rightleftarrows$1 transition at the temperature $T$. By considering the solutions at infinite storage times,
the asymptotic $J$=1 fraction is simply $F=B_{01}\rho^T(\omega)/\left(A_{10}+B_{10}\rho^T(\omega)+B_{01}\rho^T(\omega)\right)$.
Combining these two expressions and noting that $B_{01}$=3$B_{10}$ in the present case, we get
\begin{equation}
A_{10}=\Gamma_{\mathrm{eff}}\left(1\mathrm{-}4F/3\right)
\label{WolfMagic}
\end{equation}
and an intrinsic lifetime of the $J$=1 level of $\tau$=$A_{10}^{-1}$=145$\pm$28~s.
This result is consistent with a
recently calculated value of 1.10~D for the electric-dipole moment of the OH$^-$ electronic ground state~\cite{Vamhindi}, which suggests a theoretical intrinsic $J$=1 lifetime of 150~s.


\section{Conclusion and outlook}

In this Letter, we have reported measurements of the rotational temperature of OH$^-$ ions as functions of the time they are stored in an ion-beam storage ring operating at 13.5$\pm$0.5~K. In the long storage-time limit, the internal excitation distribution of the stored OH$^-$ ions is found to be well characterized by a  temperature of 13.4$\pm$0.2~K where only about one in twenty OH$^-$ ions are rotationally excited and the rest are in the $J$=0 rotational ground state.
The fraction of rotationally excited ions was further reduced to 0.9$\pm$0.1\,\% by merging  a depletion laser with the stored OH$^-$ beam on one side of the ring while probing the rotational distribution with a crossed laser beam on the other side.
After applying the depletion laser for about 400 seconds and then shutting it off, the population slowly increases to the asymptotic value of 5.1$\pm$0.3\,\% as the remaining ion ensemble again reaches thermal equilibrium with the storage ring.
Such repopulation measurements yield the effective radiative lifetime of the $J$=1 level and after corrections for the influence of the 13.4$\pm$0.2~K blackbody-radiation field, we deduce an intrinsic $J$=1 lifetime of $\tau$=$A_{10}^{-1}$=145$\pm$28~s, in agreement with the most recent theory \cite{Vamhindi}. 

While the OH$^-$ ion is particularly well-suited for the present experiment due to its comparatively large rotational constant, the cooling observed here will
occur for any
infrared-active molecular ion
regardless of its specific energy level structure. 
The present results are very promising for future experiments where the quantum state distributions of an ion ensemble needs to be strongly limited before it is brought to interact with photons~\cite{Tho11,DESIREE_Comm,CSR,CSR11,RICE,RICE12,Wolf_15}, free electrons~\cite{CSR,CSR11}, neutrals~\cite{CSR,CSR11}, or other ions~\cite{Tho11,DESIREE_Comm}. The ability to cool the ions to very low internal temperatures is also essential for action spectroscopy on cold interstellar ions.


\begin{acknowledgments}
This work was supported by the Swedish Research Council (Contracts No. 821-2013-1642, No. 621-2015-04990, No. 621-2014-4501, No. 621-2016-06625, No. 621-2016-04181) and the Knut and Alice Wallenberg Foundation. We acknowledge support from the COST Action No. CM1204 XUV/X-ray light and fast ions for ultrafast chemistry (XLIC). M.K. acknowledges financial support from the Mobility Plus Program (Project No. 1302/MOB/IV/2015/0) funded by the Polish Ministry of Science and Higher Education.
\end{acknowledgments}



\end{document}